\documentclass[prd,onecolumn]{revtex4}
\usepackage{epsfig,amsmath,amssymb,mathtools,amscd}
\usepackage[active]{srcltx}
\usepackage[linkcolor=red]{hyperref} 
\usepackage{dsfont}
\usepackage{subfigure}

\def\bvec#1{\mathbf{#1}}
\def\bx{\bvec x}
\def\bk{\bvec k}

\def\bn{\bvec n}

 \def\>{\rangle} \def\<{\langle}
 
  \def\Reals{{\mathbb R}}

\def\ket#1{| #1 \rangle} 
\def\bra#1{\langle #1 |}
\def\ketbra#1#2{| #1 \rangle \langle#2 |}

\def\d#1 {\mathop{\!\! \mathrm{d}#1}\,}
\def\df#1#2 {\!\!\frac{\mathop{\mathrm{d}#1}}{#2}\,}

\def\M{\mathcal D}
\def\B{{L}^D_\beta}
\def\L{L_\beta}

\begin{document}

\title{Weyl, Dirac and Maxwell Quantum Cellular Automata: \\
analitical solutions and phenomenological predictions of the
  Quantum Cellular Automata Theory of Free Fields\footnote{Work presented (together with Ref.\cite{bisio2015free}) at the conference \emph{Quantum
      Theory: from Problems to Advances}, held on 9-12 June 2014 at at
    Linnaeus University, Växjö University, Sweden.}}

\author{Alessandro \surname{Bisio}} \email[]{alessandro.bisio@unipv.it} \affiliation{QUIT group, Dipartimento di
  Fisica, Universit\`{a} degli Studi di Pavia, via Bassi 6, 27100 Pavia, Italy}
\affiliation{Istituto Nazionale di Fisica Nucleare, Gruppo IV, via Bassi 6, 27100 Pavia, Italy}

\author{Giacomo Mauro \surname{D'Ariano}} \email[]{dariano@unipv.it}
\affiliation{QUIT group, Dipartimento di Fisica, Universit\`{a} degli
   Studi di Pavia, via Bassi 6, 27100
  Pavia, Italy}
\affiliation{Istituto Nazionale di
  Fisica  Nucleare, Gruppo IV, via Bassi 6, 27100
  Pavia, Italy}

\author{Paolo \surname{Perinotti}}
\email[]{paolo.perinotti@unipv.it} 
\affiliation{QUIT group, Dipartimento di Fisica, Universit\`{a} degli
   Studi di Pavia, via Bassi 6, 27100
  Pavia, Italy}
\affiliation{Istituto Nazionale di
  Fisica  Nucleare, Gruppo IV, via Bassi 6, 27100
  Pavia, Italy}

\author{Alessandro \surname{Tosini}}
\email[]{paolo.perinotti@unipv.it} 
\affiliation{QUIT group, Dipartimento di Fisica, Universit\`{a} degli
   Studi di Pavia, via Bassi 6, 27100
  Pavia, Italy}
\affiliation{Istituto Nazionale di
  Fisica  Nucleare, Gruppo IV, via Bassi 6, 27100
  Pavia, Italy}

\begin{abstract}
Recent advances on quantum foundations achieved the derivation of free quantum field theory from general principles, without referring to mechanical notions and relativistic invariance. From the aforementioned principles a quantum cellular automata (QCA) theory follows, whose relativistic limit of small wave-vector provides the free dynamics of quantum field theory. The QCA theory can be regarded as an extended quantum field theory that describes in a unified way all scales ranging from an hypothetical discrete Planck scale up to the usual Fermi scale.

The present paper reviews the elementary automaton theory for the Weyl field, and the composite automata for Dirac and Maxwell fields. We then give a simple analysis of the dynamics in the momentum space in terms of a dispersive differential equation for narrowband wave-packets, and some account on the position space description in terms of a discrete path-integral approach. We then review the phenomenology of the free-field automaton and consider possible visible effects arising from the discreteness of the framework. We conclude introducing the consequences of the automaton distorted dispersion relation, leading to a deformed Lorentz covariance and to possible effects on the thermodynamics of ideal gases.  
\end{abstract}

\maketitle

\section{Introduction}

The notion of \emph{cellular automaton} was introduced by J. von Neumann in his seminal paper \cite{neumann1966theory} where he aimed at modeling a self-reproducing entity. The idea behind the concept of a cellular automaton is that the richness of states exhibited by the evolution of a macroscopic system could emerge from a simple local interaction rule among its elementary constituents. More precisely, a cellular automaton is a lattice of cells that can be in a finite number of states, together with a rule for the update of cell states from time $t$ to time $t+1$. The principal requirement for such rule is \emph{locality}: The state of the cell $\bvec{x}$ at step $t + 1$ depends on the states of a finite number of neighboring cells at step $t$. The use of classical cellular automata for simulation of quantum mechanics was proposed by 'tHooft \cite{hooft2014cellular}, followed by other authors \cite{elze2014action}.

The first author to suggest the introduction of the quantum version of cellular automata was R. Feynman in the celebrated paper of Ref.~\cite{feynman1982simulating}. Since then, the interest in \emph{quantum cellular automata}
(QCAs), has been rapidly growing, especially in the Quantum Information community, leading to many results about their general structure (see e.g Refs.\cite{schumacher2004reversible,arrighi2011unitarity,gross2012index} and references therein).
Special attention is devoted in the literature to QCAs with linear evolution, known as \emph{Quantum Walks} (QWs)
\cite{grossing1988quantum,aharonov1993quantum,ambainis2001one,reitzner2011quantum}, which were especially applied in the design of quantum algorithms 
\cite{childs2003exponential,ambainis2007quantum,magniez2007quantum,farhi2007quantum}, providing a speedup for relevant computational problems.

More recently QCAs have been considered as a new mathematical framework for Quantum Field Theory
\cite{darianovaxjo2010,darianopla,Bisio2015244,PhysRevA.90.062106,bisio2014quantum,arrighi2014dirac,arrighi2013decoupled,farrelly2014causal,farrelly2013discrete}.
Within this approach, each cell of the lattice corresponds to the
evaluation $\psi(\bvec{x})$ of a quantum field at the site $\bvec{x}$ of a lattice, with the dynamics updated in discrete time steps by a local unitary evolution.  Assuming that the lattice spacing corresponds to an hypothetical discrete Planck scale\footnote{Other approaches to discrete space-time based on p-adic numbers were studied in Refs.~\cite{albeverio2009p}.}, the usual quantum field evolution should emerge as a large scale approximation of the automaton dynamics. On the other hand, the QCA dynamics will exhibit a different behaviour at a very small scale, corresponding to ultra-relativistic wave-vectors.

The analysis of this new phenomenology is of crucial importance in providing the first step towards an experimental test of the theory as well as a valuable insight on the distinctive features of the QCA theory. Until now the research was mainly focused on linear QCAs which describe the dynamics of free field. By means of a Fourier transform the linear dynamics can be easily integrated and then, as we will show in section \ref{sec:interp-hamilt-diff},
an approximated model for the evolution of particle states (i.e. state of the dynamics narrow-band in wave-vector) can be obtained. 
Moreover, it is also possible to derive an analytical solution of the evolution in terms  of a path sum in the position space, thus giving the QCA analog of the Feynman propagator \cite{d2014path,d2014discrete}.
In section \ref{sec:phenomenology} we will exploit these tools
to explore many dynamical features of the QCA models for the Weyl and Dirac fields and to compare them with the corresponding counterpart emerging from the Weyl and Dirac Equation. We will see that, when considering massive Fermionic fields (e.g electrons) the deviations from the usual field dynamics cannot be reached by present day experiments, contrarily to the case of the QCA theory of the free electromagnetic field.
In section \ref{sec:phen-qca-theory} we will review the main
phenomenological aspects of the QCA model for free photons
(that in this framework become composite particles) with special emphasis of the emergence of a frequency-dependent speed of light, a Planck-scale effect already considered by other authors in the Quantum Gravity community
\cite{ellis1992string,lukierski1995classical,Quantidischooft1996,amelino2001testable,PhysRevLett.88.190403}.
In the final Section of this paper we address two issues of the
QCA theory that are still under investigation. The first one concerns
the notion of Lorentz covariance: Because of its intrinsic discreteness, a QCA model cannot enjoy a notion of
Lorentzian space-time and the usual Lorentz covariance must break down
at very small distances. One way of addressing the problem of changing
the reference frame 
is to assume that every inertial observer must observe the same dynamics.
Then one can look for a set of modified Lorentz transformtion which
keep the QCA dispersion relation invariant. The first step of this
analysis are reported in Section  \ref{sec:lorentz-covar-deform}.
The second issue we will briefly address in Section
\ref{sec:therm-free-ultr}
 are thermodynamical effects that could emerge from modified QCA dynamics.

\section{Weyl, Dirac and Maxwell automata}
\label{sec:weyl-dirac-maxwell}

A Quantum Cellular Automaton (QCA) describes the discrete 
time evolution of a set of cells, each one containing an array of
quantum modes. In this section we review the QCA models for the free fermions and for
the free electromagnetic field. For a complete presentation of these
results we refer to Refs. \cite{Bisio2015244,PhysRevA.90.062106,bisio2014quantum}.
  Within our
framework we will consider \emph{Fermionic} fields, our choice being
motivated by the requirement that amount of information in finite
number of cells must be finite. Then, each cell $\bvec{x}$ of the
lattice is associated with the Fermionic algebra generated by the
field operators $\{ \psi(\bvec{x}), \psi^{\dag}(\bvec{x})\}$ which
obey the canonical anticommutation relation $[\psi(\bvec{x}),
\psi^{\dag}(\bvec{x}')]_+ = \delta_{\bvec{x},\bvec{x}'}$ and
$[\psi(\bvec{x}), \psi(\bvec{x}')]_+ =0$ \footnote{We denote as
  $[A,B]_+$ the anticommutator $AB+BA$.  The commutator $AB-BA$ will
  be denoted as $[A,B]_-$.}.  With a slight genealization, we consider
the case in which each cell correspond to more than one Fermionic
mode. Different Fermionic modes will be denoted by an additional
label, e.g. $\psi_i(\bvec{x})$.  The automaton evolution will be
specified by providing the unit-step update of the Fermionic field
operators.  This rule defines the primitive physical law, and must
then be as simple and universal as possible. This principle translates
into a minimization of the amount of mathematical parameters
specifying the evolution. In particular we constrain the automaton to
describe a \emph{unitary} evolution which is \emph{linear} in the
field. We
notice that the linearity of the QCA restrict the scenario to
non-interacting field dynamics. Then we require the evolution to be
\emph{local}, which means that at each step every cell interacts with
a finite number of neighboring cells, and \emph{homogeneous}, meaning
that all the steps are the same, all the cells are identical systems
and the interactions with neigbours is the same for each cell (hence
also the number of neigbours, and the number of Fermionic modes in
each cell).  The neighboring notion also naturally defines a graph $\Gamma$
with $\bvec{x}$ as vertices and the neighboring
couples as edges.  
We also assume \emph{transitivity}, i.e. that every two cells are
connected by a path of neighbors and \emph{isotropy} which means that
the neighboring relation os symmetric and there exist a group of automorphisms for the graph under which the automaton is covariant.  From these assumptions one can show\footnote{This step
 would requires a more precise mathematical characterization (which we omit) of the 
 presented assumptions. See Ref. \cite{PhysRevA.90.062106} for the details.} that graph $\Gamma$ is a \emph{Cayley graph} of a group $G$. In the following, we consider the Abelian case $G = \mathbb{Z}^3$. 
 
Let $\mathsf{S}_+$ denote the set of generators of 
$\mathbb{Z}^3$ corresponding to the Cayley graph $\Gamma$
and let $\mathsf{S}_-$ be the set of inverse generators.
 For a given cell $\bvec{x}$ the set of neighboring cells 
is given by the set $\mathcal{N}_\bvec{x} := \{ \bvec{x} + \bvec{z} \,|\,
\bvec{z} \in  \mathsf{S} := \mathsf{S}_+ \cup \mathsf{S}_-\}$,
where we used the additive notation for the group composition.
If $s$ is the number of Fermionic modes in each cell,
the single step evolution can then be represented in terms of
$s \times s $ transition matrices $A_{\bvec{z}}$ as follows
\begin{align}
  \boldsymbol{\psi}(\bx, t+1 ) = \sum_{\bvec{z} \in \mathsf{S}} A_{\bvec{z}}
  \boldsymbol{\psi}(\bx + \bvec{z}, t).
\end{align}
where $\boldsymbol{\psi}(\bx , t)$ is the array of field operators at $\bx$
at step $t$.
Upon introducing the Hilbert space $\ell^2( \mathbb{Z}^3 )$, 
the automaton evolution can be described by the unitary matrix 
$A$ on   $\ell^2( \mathbb{Z}^3 ) \otimes \mathbb{C}^s$ given by 
 \begin{align}
A :=  \sum_{\bvec{z} \in \mathsf{S}} T_{\bvec{z}} \otimes A_{\bvec{z}},
\end{align}
where $T_\bvec{x}$ denotes the unitary representation of   
$ \mathbb{Z}^3$ on $\ell^2( \mathbb{Z}^3 )$,
$T_\bvec{y} \ket{\bx} = \ket{\bx + \bvec{y}} $.
If $s=1$ , i.e. there only one Fermionic mode in each cell,
one can prove that the only evolution which obeys our set of
assumptions is the trivial one ($A$ is the identity matrix).
Then we are led to consider the $s=2$ case and we denote the two
Fermionic modes as $\psi_L(\bx,t)$ and $\psi_R(\bx,t)$.
Moreover in the $s=2$ case one can show that our assumptions\footnote{In order to
  prove this step one need a stronger isotropy condition than the one
  presented in the text. See Ref. \cite{PhysRevA.90.062106} for the
 details.} imply that the only lattice which admits a nontrivial
evolution is the body centered cubic (BCC) one.
Being $\mathbb{Z}^3 $ an abelian group, the Fourier transform is well
defined and the operator $A$ can be block-diagonalized as follows
\begin{align}
\label{eq:autofourier}
  A = \int_B \d^3 \!\!\bk \,\, \ketbra{\bk}{\bk} \otimes A_\bk,
\end{align}
where
$|\bk\>:=(2\pi)^{-\tfrac{3}{2}}\sum_{\bx\in
  \mathbb{Z}^3}e^{i\bk\cdot\bx}|\bx\>$,
$B$ is the first Brillouin zone of the BCC lattice and 
$A_\bk:=\sum_{\bvec{z}\in S}\bvec k\,e^{-i\bvec
  k\cdot\bvec{z}}A_\bvec{z}$ is a $2 \times 2$ unitary for every
$\bk$.
We have only two (up to a local change of basis) non trivial QCAs corresponding to the unitary
matrices
\begin{equation}\label{eq:weyl3D} 
A^{\pm}_{\bk} := d^{\pm}_{\bk} I+\tilde{\bn}^{\pm}_{\bk}\cdot\boldsymbol{\sigma}
=\exp[-i\bvec{n}^{\pm}_{\bk} \cdot \boldsymbol{\sigma}],
\end{equation}
where $ \boldsymbol{\sigma}$ is the array
$(\sigma_x,\sigma_y,\sigma_z)$
of Pauli matrices and we defined
\begin{align}
&\tilde{\bn}^{\pm}_{\bk} :=
\begin{pmatrix}
s_x c_y c_z \mp c_x s_y s_z\\
\mp c_x s_y c_z - s_x c_y s_z\\
c_x c_y s_z \mp s_x s_y c_z
\end{pmatrix}\!\!,\,
{\bn}^{\pm}_{\bk}:=\frac{\lambda^{\pm}_{\bk}\tilde{\bn}^{\pm}_{\bk}}{\sin\lambda^{\pm}_{\bk}},\nonumber\\
&\d^{\pm}_{\bk} :=  (c_x c_y c_z \pm s_x s_y s_z ),\;
\lambda^{\pm}_{\bk}:=\arccos(d^{\pm}_{\bk}),\nonumber\\
&c_\alpha := \cos({k}_\alpha/\sqrt{3}),\;s_\alpha:= \sin({k}_\alpha/\sqrt{3}),\;\alpha = x,y,z.\nonumber
\end{align}
The matrices $A_{\bk}^\pm$ in Eq. \eqref{eq:weyl3D} describe the evolution of a two-component
Fermionic field,
\begin{align}
  {\psi} ({\bk},t+1) = 
A_{\bk}^\pm {\psi} ({\bk},t),
\quad
  {\psi} ({\bk},t) : =
\begin{pmatrix}   
 {\psi}_R ({\bk},t)\\
 {\psi}_L ({\bk},t)
  \end{pmatrix}.
  \end{align}
  The adimensional framework of the automaton corresponds to measure
  everything in Planck units. In such a case the limit $|{\bk}|\ll 1$
  corresponds to the relativistic limit, where on has
\begin{equation}
\bn^{\pm}({\bk})\sim\tfrac{{\bk}}{\sqrt{3}},\quad A^{\pm}_{\bk}\sim\exp[-i\tfrac{{\bk}}{\sqrt{3}} \cdot\boldsymbol{\sigma}],
\end{equation}
corresponding to the Weyl's evolution, with the
rescaling  $\tfrac{{\bk}}{\sqrt{3}} \to \bk$.
Since the QCAs $A^{+}$ and $A^{-}$ reproduce the dynamics of the
Weyl equation in the  limit $|{\bk}|\ll 1$, we refer to them as
\emph{Weyl automata}. For sake of simplicity,
in the following we will consider only one Weyl automaton, i.e. we
define $A_{\bk} := A_{\bk}^-$ and we similarly drop all the others
$\pm$ superscripts.
 This choice is completely painless since
all the methods that we will use can be easily adapted to the choice
$A_{\bk} = A_{\bk}^-$. However the two automata, beside giving the
Weyl equation for small $\bk$, exhibit a different behaviour at high
$\bk$ and we will point out those differences whenever it will be relevant.

The derivation that we sketch previously can be carried on also in the 
two dimensional case (considering QCA on Cayley graphs  of
$\mathbb{Z}^2$)
 and in the one dimensional case (considering QCA on Cayley graphs  of
$\mathbb{Z}$).
In the 2-dimensional case we obtain a unique (up to a local change of basis) the QCA on the square latticeand it leads to 
\begin{align}
\label{eq:weyl2D} 
   A^{(2D)}_{\bk}=I d^A_\bk-i\boldsymbol\sigma\cdot\bvec a^A_\bk,
\end{align}
where the functions $\bvec a_\bk$ and $d_\bk$ are expressed in terms of $k_x:=\frac{k_1+k_2}{\sqrt2}$ and $k_y:=\frac{k_1-k_2}{\sqrt2}$ as
$
 (a^A_\bk)_x:= s_xc_y
$,
$
  (a^A_\bk)_y:=c_xs_y
$,
$
  (a^A_\bk)_z:=s_xs_y
$,
$
  d^A_\bk:=c_xc_y.
$
where $c_i =\cos\tfrac{k_i}{\sqrt2}$ and $s_i=\sin\tfrac{k_i}{\sqrt2}$.
In the one dimensional case we find
\begin{align}
\label{eq:weyl1D} 
   A^{(1D)}_{k}= 
   \begin{pmatrix}
     e^{-i k} & 0 \\
      0&  e^{i k}  \\
   \end{pmatrix}
\end{align}
Both in the 2-dimensional and 1-dimensional cases the limit 
$|{\bk}|\ll 1$ gives the 2-dimensional and 1-dimensional Weyl equation
respectively (in the 2-dimensional we need the rescaling  with the
rescaling  $\tfrac{{\bk}}{\sqrt{2}} \to \bk$.).
The QCA in Eqs. (\ref{eq:weyl3D},~(\ref{eq:weyl2D}) and~(\ref{eq:weyl1D}))
describe the dynamics of free massless Fermionic fields. If we couple 
two Weyl automata $A_\bk$ with a mass term we obtain a new QCA $U_\bk$ given by
\begin{align} \label{eq:diracauto}
  U_\bk = \begin{pmatrix}
     n A_\bk & im I \\
      im I & n A_\bk^\dagger  \\
   \end{pmatrix}
\quad n^2 + m^2 =1.
\end{align}
Clearly this construction can be done in the $1$,$2$ and
$3$-dimensional cases and the resulting QCA is alway unitary and local.
One can easily see that in the limit $|{\bk}|\ll 1$  and $m \ll 1$
Eq. \ref{eq:diracauto}  (with the appropriate rescaling of $\bk$ in
$2$ and $3$ dimensions) gives the same evolution as Dirac equation
and then we denote the automata of Eq. (\ref{eq:diracauto})
\emph{Dirac automata}.

The $3$-dimensional Weyl QCA can also be use as a building block for
QCA model of free electrodynamics. The basic idea is to interpret the
photon as a pair of Weyl fermions that are suitably correlated in
wave-vector. Then one can show that, in an appropriate regime, this field
obeys the dynamics dictated by the Maxwell equations and the bosonic
commutation relation are recovered. This approach recall the so-called
beutrino theory of light of De Broglie
\cite{de1934nouvelle,jordan1935neutrinotheorie,kronig1936relativistically,perkins1972statistics,perkins2002quasibosons} which suggested that
the photon could be a composite particle made of of a
neutrino-antineutrino pair.  Within our framework (we omit the details
of this construction that can be found in
Ref. \cite{bisio2014quantum}) the electric and magnetic field are
given by
\begin{align}
  \label{eq:electric and magnetic field}
  &\bvec{E}:=|{\bn}_{\tfrac\bk2}|(\bvec{F}_T+\bvec{F}_T^\dag),\quad\bvec{B}:=i|{\bn}_{\tfrac\bk2}|(\bvec{F}_T^\dag-\bvec{F}_T),\\
  &2|{\bn}_{\tfrac\bk2}|\bvec{F}_T=\bvec{E} + i \bvec{B}
 \nonumber\\
&\bvec{F}_T(\bk) := \bvec{F}(\bk)  - 
\left(\frac{\bvec{n}_{\frac{\bk}{2}}}{|\bvec{n}_{\frac{\bk}{2}}|} \cdot
{\bvec{F}}(\bk) \right)
\frac{\bvec{n}_{\frac{\bk}{2}}}{|\bvec{n}_{\frac{\bk}{2}}|} 
\nonumber\\
&\bvec{F}(\bk) := 
(
F^{1}(\bk), F^{2}(\bk), F^{3}(\bk)
) ^T, \nonumber\\
&F^{j}(\bk) := 
 \int \frac{ d \bvec{q}}{(2 \pi)^3}
f_{\bk}(\bvec{q})
\phi
\left(\tfrac{\bk}{2}-\bvec{q}\right)
\sigma^{j}
\psi
 \left(\tfrac{\bk}{2}+\bvec{q}\right) \nonumber
\end{align}
where
$\int\frac{d\bvec{q}}{(2\pi)^3} |f_{\bk}(\bvec{q})|^2 =1, \forall \bk$
and $\phi
\left(\bk  \right)$, $\psi
\left(\bk  \right)$  
are two massless Fermionic fields whose evolution is dictated by the
automaton
$A_\bk^*$ and $A_\bk$ respectively\footnote{We denote as $A^*$ the
  complex conjugate of $A$}, i.e.
\begin{align}
  \phi
\left(\bk , t \right) &= 
A_\bk^{* t}  \phi
\left(\bk   \right)  
\qquad
  \psi
\left(\bk , t \right) = 
A_\bk^{t}  \phi
\left(\bk   \right)  \nonumber \\
  \phi
\left(\bk \right) &= 
\begin{pmatrix}
   \phi_R
\left(\bk \right)  \\
\phi_L
\left(\bk \right)
\end{pmatrix}
\qquad
  \psi
\left(\bk \right) = 
\begin{pmatrix}
   \psi_R
\left(\bk \right)  \\
\psi_L
\left(\bk \right)
\end{pmatrix} 
\nonumber .
\end{align}
For an appropriate choice of the functions $f_{\bk}(\bvec{q})$ (see
Ref. \cite{bisio2014quantum}), one can prove that the evolution of the
Electric and Magnetic fields which are defined in
Eq.\eqref{eq:electric and magnetic field} is given by the following
equations\footnote{Since a QCA describe an evolution which discrete in
  time, the derivative with respect time is not defined in this
  context. However we can imagine 
$A_\bk^{t}$ to be defined for any
  real value of $t$ and then derive with respect the continuous
  variable $t$. This is the construction which underlies Eq.~(\ref{eq:maxwell2})}
\begin{align}
  \begin{split} \label{eq:maxwell2}
    &\partial_t\bvec{F}_T(\bk,t) =
2\bvec{n}_{\frac{\bk}{2}} \times \bvec{F}_T(\bk,t)\\
&2\bvec{n}_{\frac{\bk}{2}} \cdot \bvec{F}_T(\bk,t) = 0,
  \end{split}
\end{align}
which in the limits $|{\bk}|\ll 1$ and with the rescaling
$\tfrac{\bk}{\sqrt{3}} \to \bk$
 become the Fourier trasform of the usual vacuum Maxwell equations
in position space
\begin{align}
      \label{eq:maxwellstandard}
  \begin{array}{lcl}
    \nabla \cdot \bvec{E}=0    & &\nabla \cdot \bvec{B} =0\\
    \partial_t \bvec{E} = c\nabla \times \bvec{B}  && \partial_t \bvec{B} = -c\nabla \times \bvec{E} \;\;.
  \end{array}
\end{align}
Because of this result, we refer to the construcion of Eq. (\ref{eq:electric
  and magnetic field}) as the \emph{Maxwell automaton}. This is a
slight abuse of notation since Eq. (\ref{eq:electric and magnetic
  field}) does not introduce any new QCA model but it defines a field
of bilinear operators, each one of them evolving with the Weyl
automaton, that  can be interpreted as the electromagnetic field in
vacuum. In this sense the expression ``Maxwell automaton'' actually
means QCA model for the Maxwell equations (in vacuum).

\section{Analysis of the dynamics}

The aim of this section is to analyse the dynamics of the QCA models
presented in the previous section.  The material of this section can
be found in Refs.
\cite{Bisio2015244,PhysRevA.90.062106,bisio2013dirac,d2014path,d2014discrete}.

In this section we fosus on the single particle sector of the QCA. Since the
QCAs we are considering are linear in the fields the single particle
sector contains all the information of the dynamics (it is a free
theory). We can then write $\ket{\psi (t)} = A^t \ket{\psi(0)}$ where 
$\ket{\psi (t)} := \sum_x \ket{\psi(x,t)} \ket{x}$ is generic  one particle state. This kind of framework is
better known in the literature under the name \emph{Quantum Walk}
\cite{grossing1988quantum,aharonov1993quantum,ambainis2001one,succi1993lattice,bialynicki1994weyl,meyer1996quantum}.

\subsection{Interpolating Hamiltonian and differential equation for single-particle wave-packets} \label{sec:interp-hamilt-diff}

Since a QCA (and a Quantum Walk) describes a discrete evolution on a lattice,
the notion of Hamiltonian (like any other differential operators)
is completely deprived of physical meaning.
However it is useful to introduce a Hamiltonian operator $H_\bk$, that we
call \emph{interpolating Hamiltonian} that obeys the following equation
\begin{align}
\label{eq:interpolhamil}
  A^t_\bk = e^{it H_\bk}.
\end{align}
where $A_\bk^{t}$ is defined for any real value of $t$ (the automaton
$A_\bk$ can be any of the QCA models of Section
\ref{sec:weyl-dirac-maxwell}).  It is clear that $ H_\bk$ is the
genrator of the continuous time evolution which interpolate the QCA
dynamics between the integer steps.  The eigenvalue of $H_\bk$ have
the same modulus and its analytical expression, denoted by
$\omega(\bk)$, is the \emph{dispersive relation} of the automaton and
it provides a lot of information about the dynamics.  In analogy with
what we do in Quantum Field Theory, states of the dynamics
corresponding to the positive eigenvalue $\omega(\bk)$ are called
\emph{particle states} while eigenstate with negative eigenvalue
$-\omega(\bk)$ are called \emph{antiparticle states}.
If we denote with $\ket{u}_\bk$ a positive frequency eigenstate of 
 $H_\bk$ (i.e. $H_\bk  \ket{u}_\bk = \omega(\bk) \ket{u}_\bk$)
a generic particle state with positive frequency
\begin{align}\label{eq:particlestates}
\ket{\psi}_+ =  \int  \frac{d\bvec{k}}{(2\pi)^n}  g(\bk)
\ket{u}_\bk \ket{\bk}
\end{align}
where $n$ is the dimension of the lattice and $g(\bk)$ is a normalized
probability amplitude. We remind that for the $2$ and $3$-dimensional
Dirac QCA $H_\bk$ has dimension $4$ and then both the eigenvalues 
have degeneracy (corresponding to the spin degree of freedom).
The construction of Eq. (\ref{eq:particlestates}) can be
straightforwardly applied also for the definition general antiparticle
states with negative frequency $\ket{\psi}_-$.

\begin{figure}[h!]
  \centering
\includegraphics[width=0.45\textwidth]{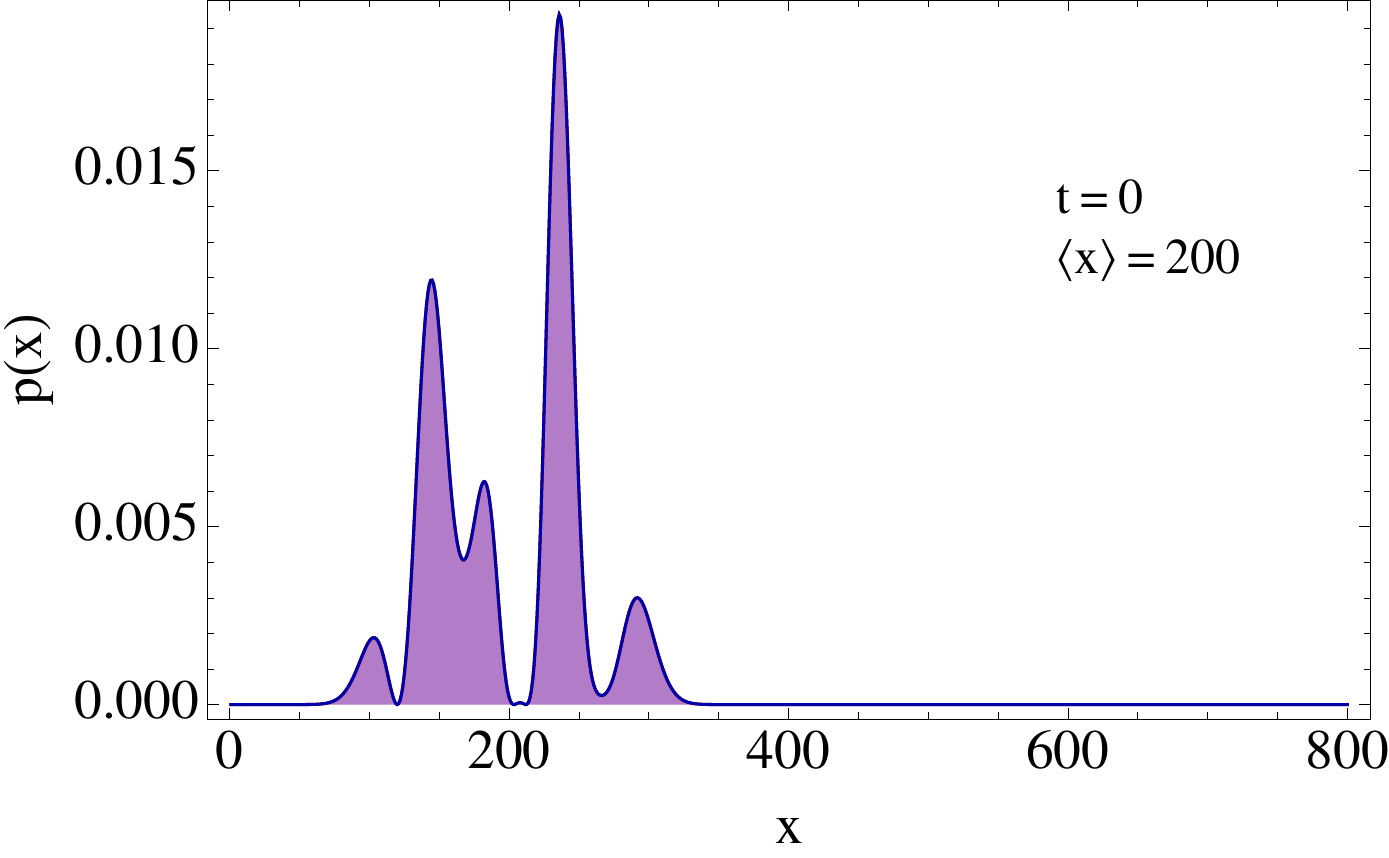}
\qquad 
\includegraphics[width=0.45\textwidth]{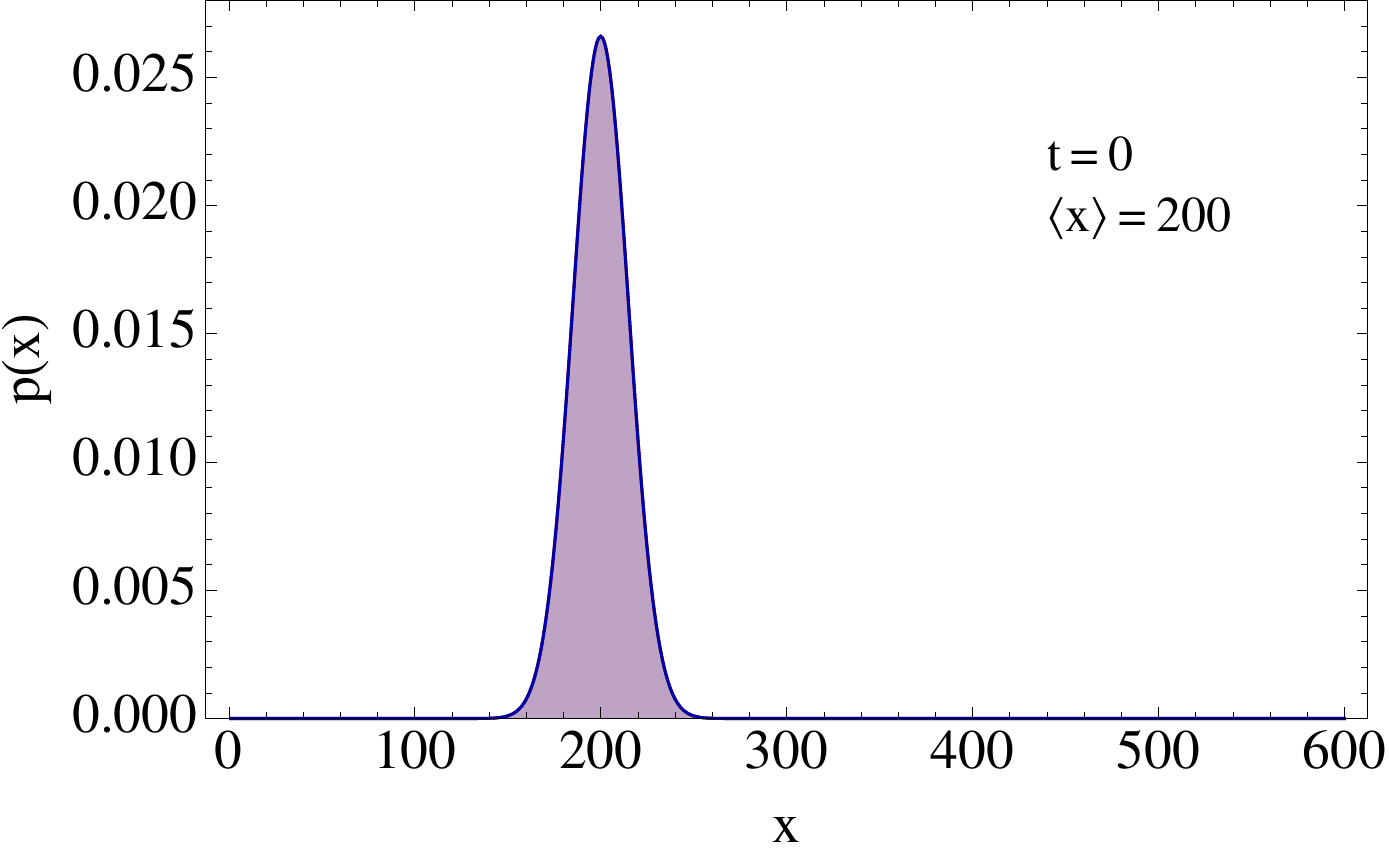}\\
\includegraphics[width=0.45\textwidth]{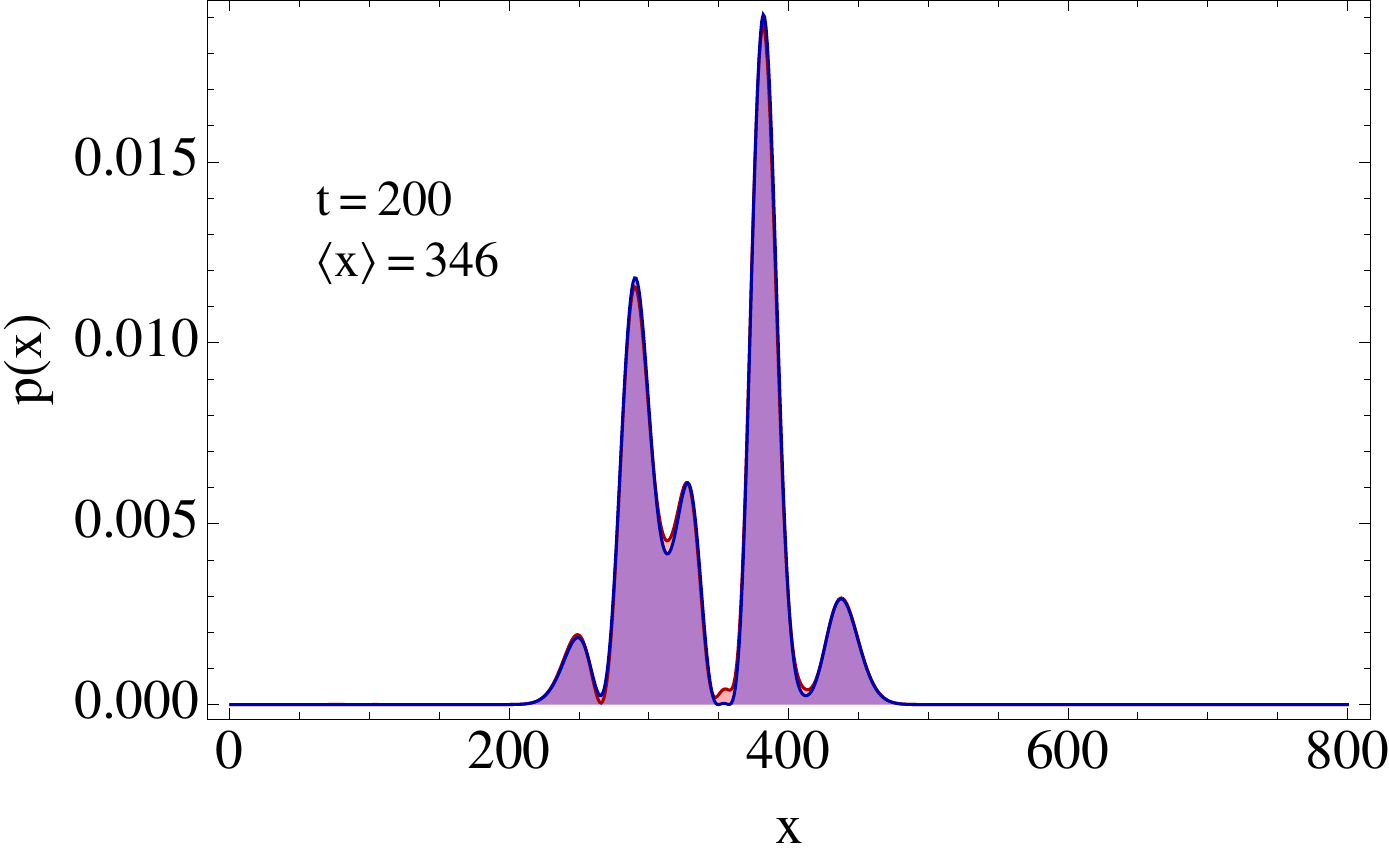}
\qquad 
\includegraphics[width=0.45\textwidth]{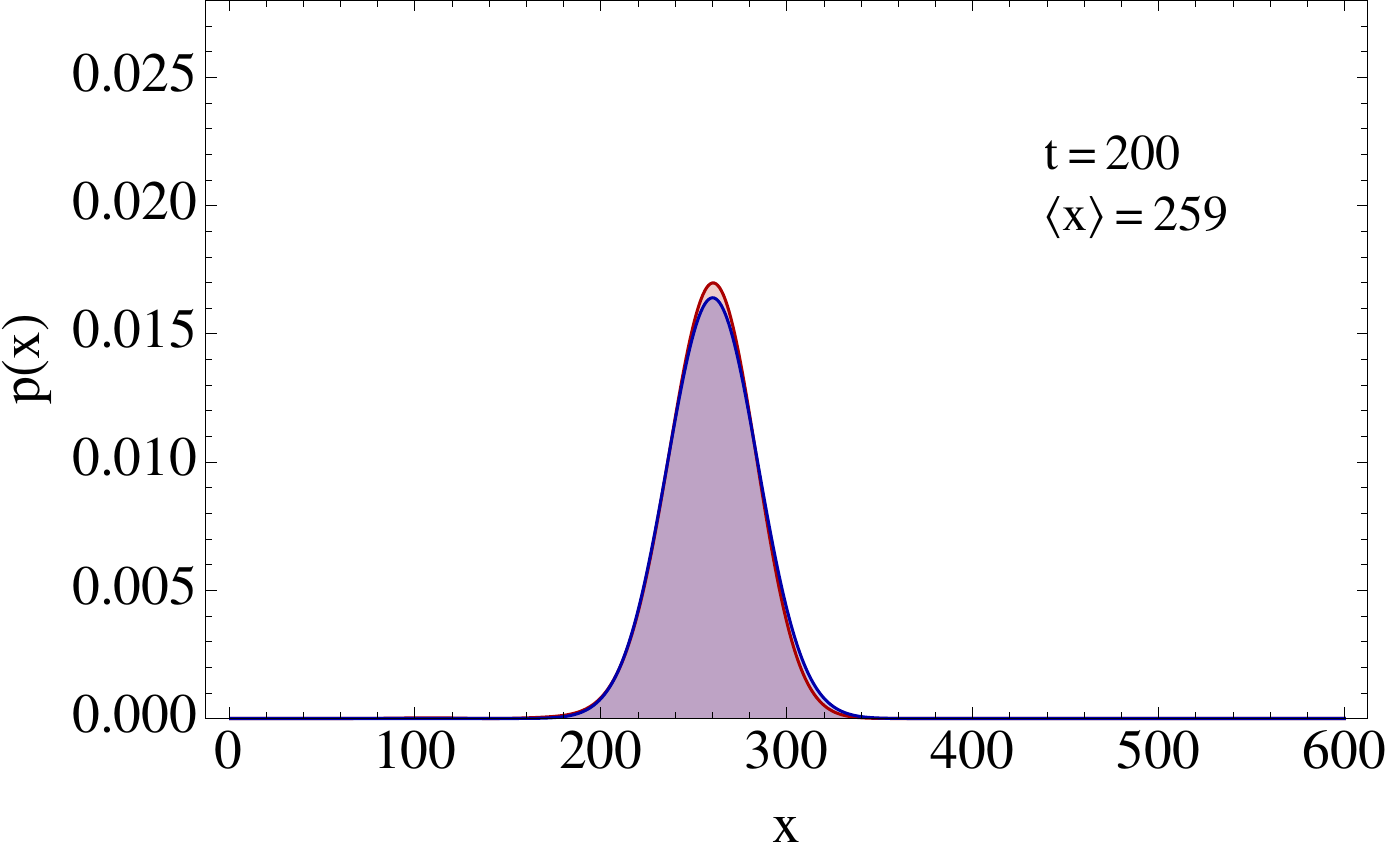}\\
\includegraphics[width=0.45\textwidth]{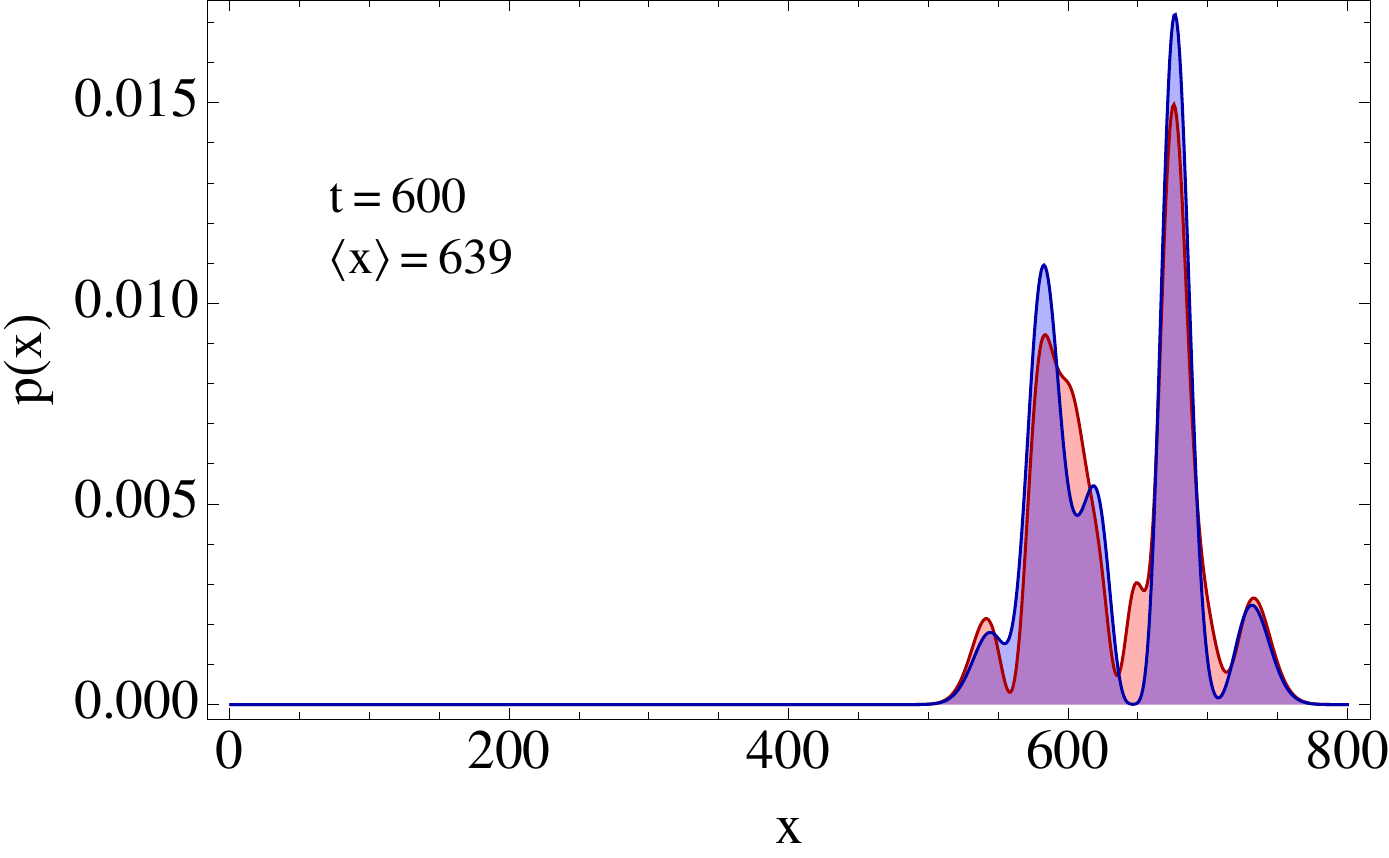}
\qquad 
\includegraphics[width=0.45\textwidth]{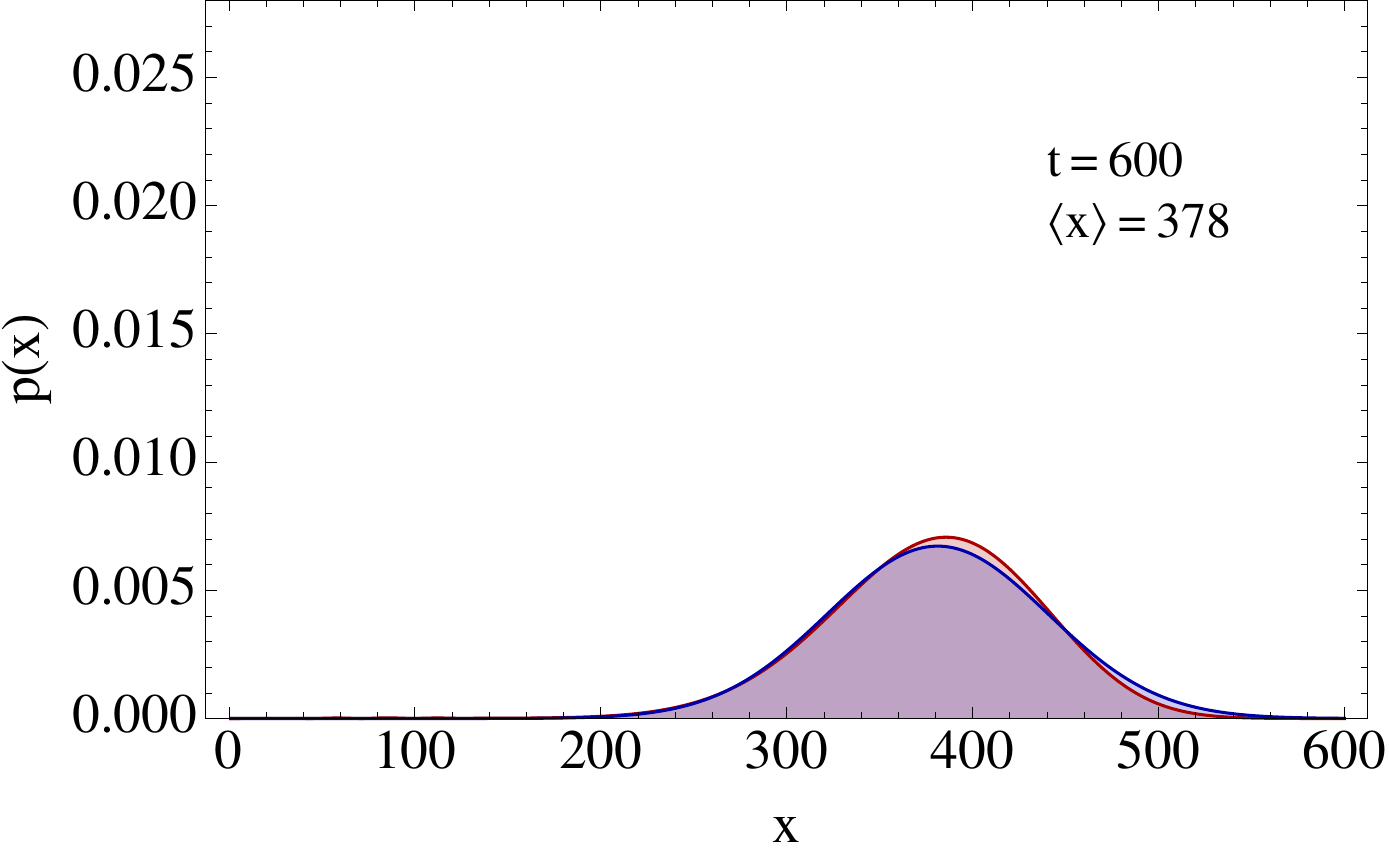}
\caption{(Colors online) Test of the approximated evolution of Eq.
\eqref{eq:schroding2} of the one dimensional Dirac automaton evolution. {\bf Left figure:} here the
  state is a superposition of Hermite functions
  (the polynomials $H_j(x)$ multiplied by the Gaussian) peaked around $k_0=3\pi/10$. {\bf Right
    figure:} here the initial state is Gaussian profile peaked around
  $k_0=0.1$. This figure is published in Ref. \cite{Bisio2015244}.}\label{fig:Hermite}
\end{figure}

One can use the
interpolating Hamiltonian $H_\bk$ in order to rephrase the continuous
evolution in terms of a differential Equation, i.e.
 \begin{align}
\label{eq:schroding1}
  i \partial_t \ket{\psi(\bk,t)} =  H_\bk \ket{\psi(\bk,t)}.
 \end{align}
where $\ket{\psi(\bk,t)}$ is the wave-vector representation of a
one-particle state, i.e. $\ket{\psi} = \int  \tfrac{d\bvec{k}}{(2\pi)^3}   \ket{\psi(\bk,t)}
\ket{\bk}$ .
When  the initial state has positive frequency (see
 Eq. (\ref{eq:particlestates})) and its distribution  $g(\bk)$ is
 smoothly peaked around a given $\bk_0$, the evolution of Eq.~(\ref{eq:schroding1})
can be approximated by the following dispersive equation 
\begin{equation}
\label{eq:schroding2}
i\partial_t \tilde{g}(\bvec x,t)=\pm[\bvec v\cdot\bvec \nabla+\tfrac{1}{2}\bvec D\cdot\bvec
\nabla\bvec \nabla]\tilde{g}(\bvec x,t),
\end{equation}
where $\tilde{g}(\bvec x,t)$ is the Fourier transform of
$\tilde{g}(\bk,t):=e^{-i\bk_0\cdot\bvec x+i\omega(\bk_0) t}\psi(\bk,t)$, and $\bvec v$ and $\bvec D$ are the drift
vector $\bvec v=\left(\bvec\nabla_{\bk}\omega\right)(\bk_0)$ and diffusion tensor
$\bvec
D=\left(\bvec\nabla_{\bk}\bvec\nabla_{\bk}\omega\right)(\bk_0)$,
respectively.
Intuitively the vector $\bvec v$ represent the velocity of the
wavepacket and the tensor $\bvec D$ tells us how the
wavepacket spreads during the evolution.
The accuracy of the approximation can be analytical evaluated (see
Ref. \cite{Bisio2015244}) and compared with comuter simulation as in Fig.~\ref{fig:Hermite}.

\section{Phenomenology}
\label{sec:phenomenology}
This section is devoted to the study of the various phenomenological
effects of the QCA model presented in Section
\ref{sec:weyl-dirac-maxwell}. The aim of this analysis is to
understand the properties of the QCA dynamics and compare its features
to the known results about the dynamics of free quantum fields. The
ultimate goal to identify experimental situation in which it is
possible to falsify the validity of the QCA theory.
 
\subsection{Zitterbewegung} 
The first feature of the QCA dynamics we are going to explore (for a
more complete presentation see Ref.\cite{bisio2013dirac}) is the
appearence of a fluctuation of the position in the particle
trajectory, the so called \emph{zitterbewegung}.
 
The Zitterbewegung was first recognized by Schr\"odinger in 1930
\cite{schrodinger1930kraftefreie} who noticed that in the Dirac
equation describing the free relativistic electron the velocity
operator does not commute with the Dirac Hamiltonian: the evolution of
the position operator,exhibits a
very fast periodic oscillation around the mean position with frequency $2mc^2$ and amplitude
equal to the Compton wavelength $\hbar/mc$ with $m$ the rest mass of
the relativistic particle.  Zitterbewegung oscillations cannot be
directly observed by current experimental techniques for an
electron since the amplitude is very small $\approx 10^{-12}$
m.  However, it can be seen in a number of solid-state,
atomic-physics, photonic-cristal and optical waveguide simulators 
\cite{lurie1970zitterbewegung,cannata1991effects,ferrari1990nonrelativistic,cannata1990dirac,PhysRevLett.100.113903}.

\begin{figure}[t!]
\includegraphics[width=.23\textwidth]{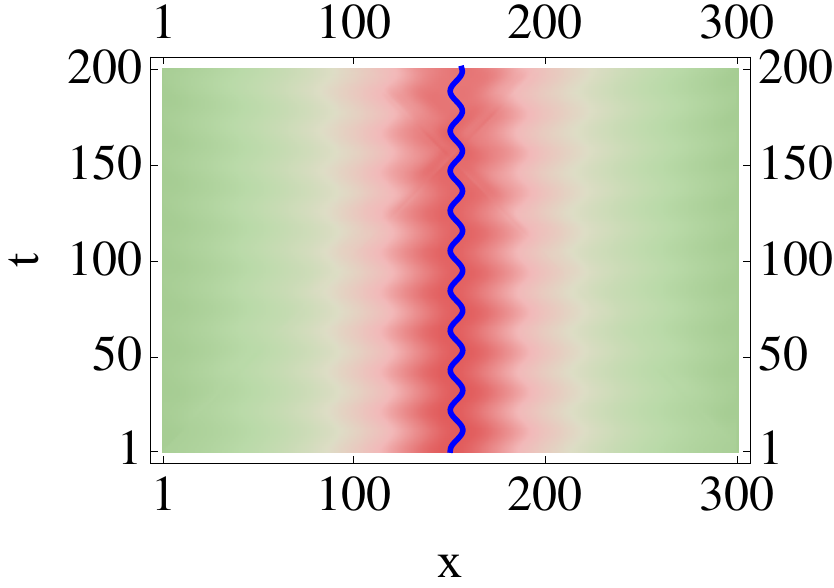}\quad
\includegraphics[width=.23\textwidth]{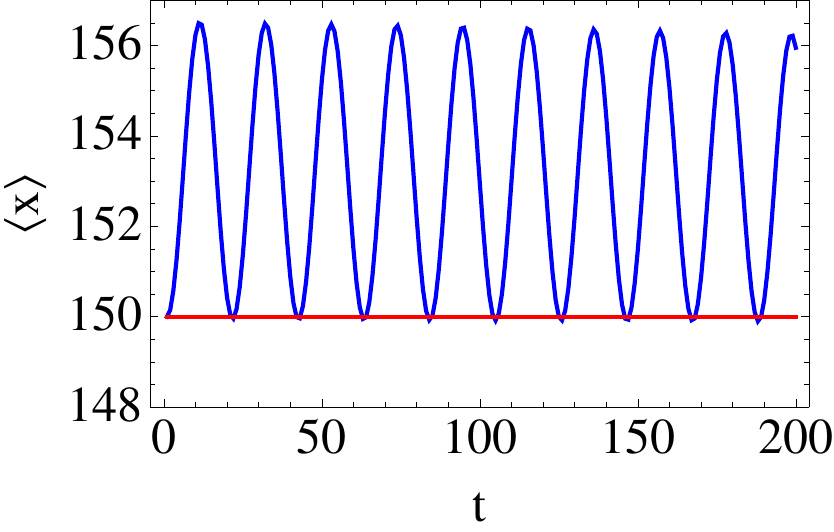}
\includegraphics[width=.23\textwidth]{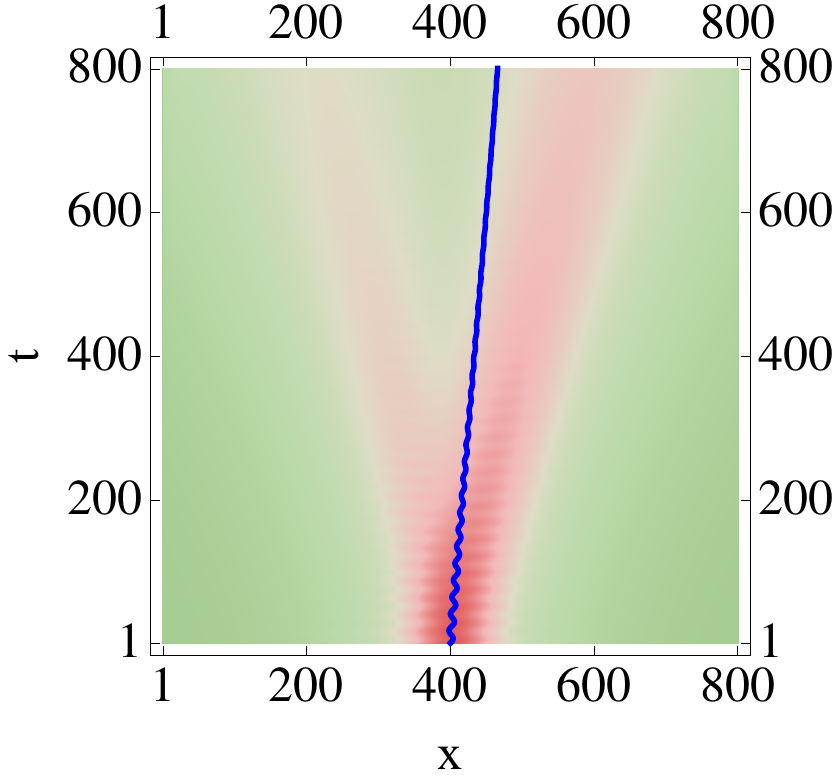}\quad
\includegraphics[width=.23\textwidth]{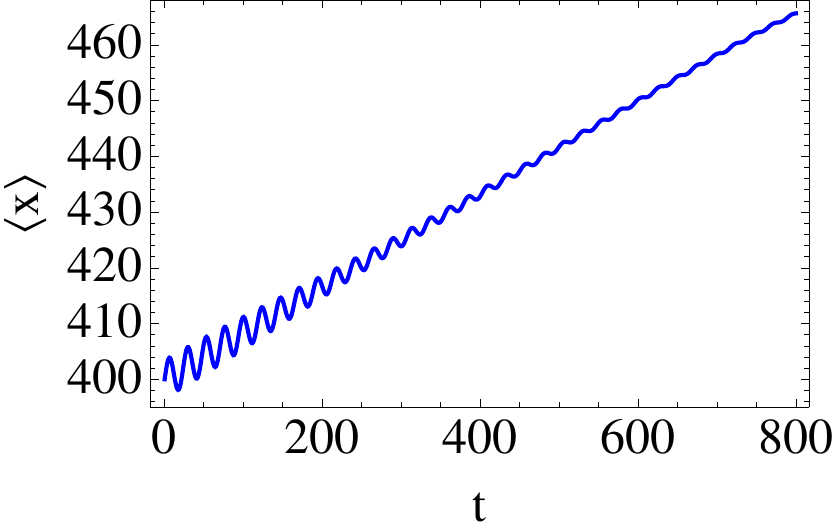}
\caption{ Zitterbewegung in the one dimensiona Dirac QCA.  {\bf
    Top:} The mass of the particle is $m=0.15$. The amplitudes of the
superposition between
  positive and negative frequency states are $c_+=1/ \sqrt{2}$
  $c_-=i/\sqrt{2}$ respectively. The wavepacket is peaked around
  $k_0=0$.
 The shift and oscillation frequency are
  respectively $\bra{\psi} X(0)+ Z_{X}(0)\ket{\psi}=3.2$ and
  $\omega(0)/\pi=0.05$.  {\bf Middle:}
  $m=0.15$, $c_+=1/\sqrt{2}$, $c_-=1/\sqrt{2}$, $k_0=0$,
  $\sigma=40^{-1}$.  The shift and oscillation frequency
  are $0$ and $0.13$, respectively.  {\bf Bottom:} $m=0.13$,
  $c_+=\sqrt{2/3}$, $c_-=1/\sqrt{3}$, $k_0=10^{-2}\pi$,
  $\sigma=40^{-1}$.  In this case the particle and antiparticle
  contribution are not balanced and the average position drift
  velocity is thus $\bra{\psi _+} V \ket{\psi_+}+\bra{\psi _-} V
  \ket{\psi_-}=(|c_+|^2-|c_-|^2)v(k_0)=0.08$, corresponding to an
  average position $x_{\psi}^+(800) + x_{\psi}^-(800)=464$. 
Notice that for $t \to \infty$ the
  term $2 \Re [\bra{\psi_+} {Z}_{{X}}(t) \ket{\psi_-}$, which is
  responsible of the oscillation, goes to
  $0$. This figure is published in Ref. \cite{bisio2013dirac}.}\label{fig:Zitt11-12-21-22}
\end{figure}

Here we focus on the one-dimensional Dirac QCA 
whose epression,  introduced in Section
\ref{sec:weyl-dirac-maxwell}, is easily obtained as special case 
of Eq.~\eqref{eq:diracauto}\footnote{More precisely,
  Eq.~\eqref{eq:weyl1D} leads to two identical copies of Eq.~\eqref{eq:Dirac1d}}
\begin{align}
  \label{eq:Dirac1d}
  U = \int_{-\pi}^{\pi} dk \ketbra{k}{k} \otimes U_k \quad 
U_k = 
\begin{pmatrix}
  n e^{-ik}  & im \\
  im  &n e^{ik} 
\end{pmatrix}
\end{align}
The ``position'' operator $X$ corresponding to the representation $|x\>$ (i.e. such that
$X|s\>|x\>=x|s\>|x\>$, $x \in \mathbb{Z}$) is defined as follows
\begin{align}
X=\sum_{x\in\mathbb{Z}}x(I\otimes\ketbra{x}{x}).
\end{align}
and it provides the average location of a wavepacket in terms
of $\<\psi|X|\psi\>$.
If we write the single particle 
in terms of its positive frequency and negative frequency components, i.e.
$|\psi\> = c_+ |\psi\>_+  +  c_- |\psi\>_-$,
yhe time evolution of the mean value of the position operator
$\bra{\psi}X(t)\ket{\psi}$ is given by
\begin{align}\nonumber
x_\psi(t) :=  \bra{\psi} X(t) \ket{\psi} =
x_{\psi}^+(t) + x_{\psi}^-(t) + x_{\psi}^{\rm{int}}(t)\\\nonumber
x_{\psi}^{\pm}(t) : = \bra{\psi _\pm} X(0) + Vt \ket{\psi_\pm}\\\label{e:zitterbewegung}
x_{\psi}^{\rm{int}}(t) :=  
2 \Re [\bra{\psi_+} X(0) - {Z}_{{X}}(0) + {Z}_{{X}}(t) \ket{\psi_-}]
\end{align}
where $V$ is a time independent operator corresponding to the group
velocity and $Z_X(t)$ is the operator that gives the oscillatory
motion (see Ref. \cite{bisio2013dirac} for the details).
We notice that the interference between positive
and negative frequency is responsible of the oscillating term
$x_{\psi}^{\rm{int}}(t)$ whose magnitude is bounded by $1/m$ which in the usual dimensional units
corresponds to the Compton wavelength $\hbar/ m c$.  These results
show that $x_{\psi}^{\rm{int}}(t)$ is the automaton analogue of the
 Zitterbewegung for a Dirac particle.  for $t \to \infty$ the term $2 \Re [\bra{\psi_+} {Z}_{{X}}(t)
\ket{\psi_-}]$, which is responsible of the oscillation, goes to $0$
as $1/\sqrt{t}$ and only the additional shift contribution given by $2 \Re
[\bra{\psi_+} X(0) - {Z}_{{X}}(0) \ket{\psi_-}]$ survives.
In Fig. \ref{fig:Zitt11-12-21-22} one can se the simulation of the evolution of states with particle
and antiparticle components smoothly peaked around some $k_0$.

\subsection{Scattering against a potential barrier}\label{s:barrier}
In this section we study the dynamics of the one dimensional Dirac
automaton in the presence of a potential.
In the position representation the one particle evolution of the 
one dimensioanal Dirac QCA
reads as follows:
\begin{align}
\label{eq:dirac1dposition}
  U := \sum_x 
\left(  
\begin{array}{ll}
   n \ketbra{x-1}{x} &   -im \ketbra{x}{x} \\
  -im \ketbra{x}{x} &   n \ketbra{x+1}{x} 
  \end{array}
\right).
\end{align}
The presence of a  potential $\phi(x)$, modifies the unitary evolution
of Eq. \eqref{eq:dirac1dposition} with a position dependent phase as
follows (see also Ref \cite{kurzynski2008relativistic,meyer1997quantum}):
\begin{align*}
  U_\phi := \sum_x e^{-i \phi(x)}
\left(  
\begin{array}{ll}
   n \ketbra{x-1}{x} &   -im \ketbra{x}{x} \\
  -im \ketbra{x}{x} &   n \ketbra{x+1}{x} 
  \end{array}
\right).
\end{align*}
We now review the analysis (carried on in Ref. \cite{bisio2013dirac}) of the case in which
 $\phi(x) := \phi \, \theta(x)$
($\theta(x)$ is the Heaviside step function) that is a potential step  which is $0$
for $x <0 $ and has a constant value $\phi \in
[0, 2\pi]$ for
$x \geq 0$.
Let us consider the situation in which, for $t \ll 0$, the state is 
a positive frequency
wavepacket peaked around $\bk_0$ that moves at group velocity
$v(k_0)$ and hits the barrier form the left.
Then and one can show that for $t \gg 0$ the state is evolved into a  superposition of a reflected
and a transmitted wavepacket as follows (we use the notation of
Eq.~\eqref{eq:particlestates} adapted at the one-dimensional case):
\begin{align*}
  \begin{split}
    \ket{\psi(t)} \xrightarrow{ t \gg 0} \beta(k_0) \int \df{k}{\sqrt{2\pi}}  g_{k_0}(k) e^{-i \omega(k)t}  \ket{u}_{-k}\ket{k} +\\
 {}+ \tilde{\gamma}(k_0) e^{-i \phi t} \int \df{k}{\sqrt{2\pi}}  \tilde{g}_{k'_0}(k') e^{-i
    \omega(k')t} \ket{u}_{k'}\ket{k'} 
  \end{split}
\end{align*}
where we defined
\begin{align*}
&k'_0 \mbox{ s.t. }    \omega(k'_0) = \omega(k_0) -\phi, \\
&\tilde{\gamma}(k_0) := {\gamma}(k_0)
\sqrt{\frac{v(k'_0)}{v(k_0)}}, \qquad
\tilde{g}_{k'_0}(k') =
\sqrt{\frac{v(k'_0)}{v(k_0)}}{g}_{k'_0}(k') 
\end{align*}
(one can check $ \int \df{k}{\sqrt{2\pi}} 
|\tilde{g}_{k'_0}(k')|^2 = 1$), whose group velocities are
$-v(k_0)$ for the reflected wave packet and $v(k^\prime_0)$ for the
transmitted wave packet.

The probability of finding the particle in the reflected wavepacket is
then $R = |\beta(k_0)|^2$
(reflection coefficient) while the probability of finding the particle in the transmitted
wavepacket is $ T= |\tilde{\gamma}(k_0)|^2$ (trasmission coefficient). The consistency of the result
can be verified by checking that $R+T = 1$.
Clearly $\phi=0$ implies
$R = 0$ and increasing $\phi$ for a fixed $k$ increases the value of
$R$ up to $R=1$.  
By further increasing $\phi$ a
transmitted wave reappears and the reflection coefficient decreases. This is the so called ``Klein
paradox'' which is originated by the presence of positive and negative frequency eigenvalues of the
unitary evolution.  The width of the $R=1$ region is an increasing function of the mass equal to
$2\arccos(n)$ which is the gap between positive and negative frequency solutions.
\begin{figure}[t!]
\includegraphics[width=.4\textwidth]{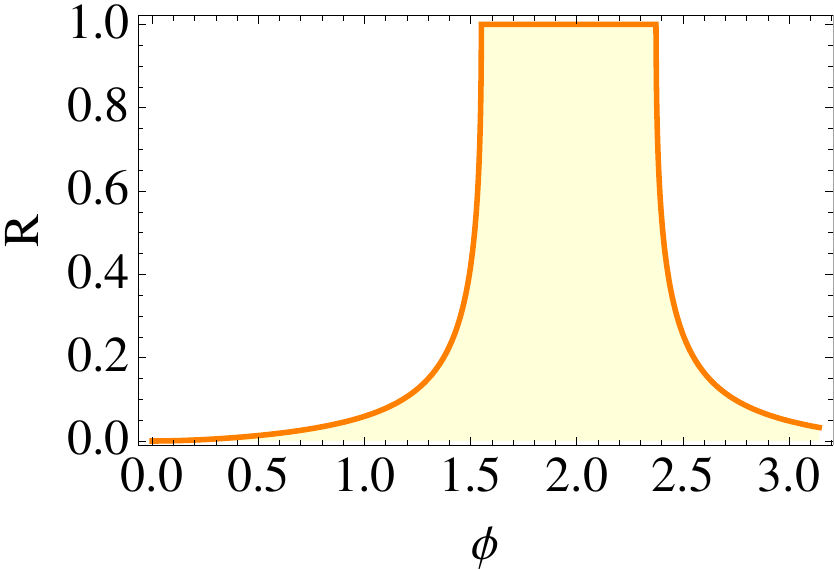}\qquad
\includegraphics[width=.4\textwidth]{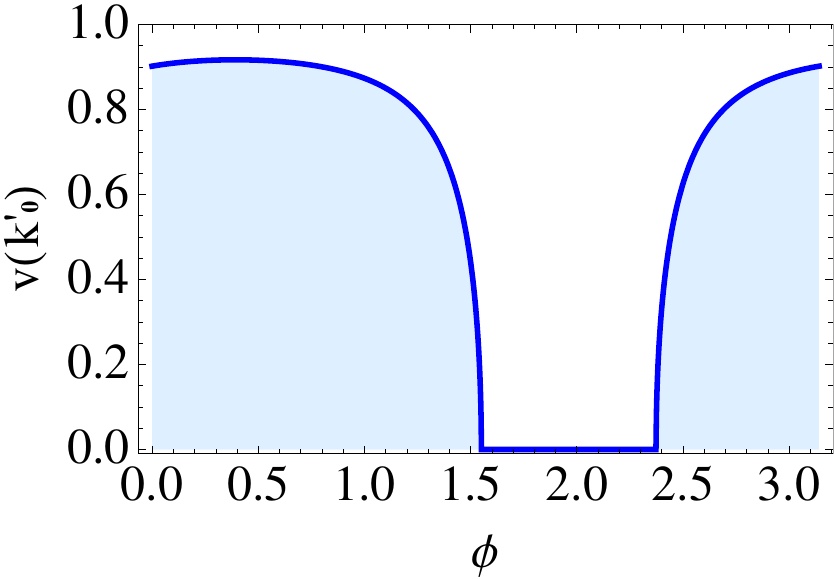}
\caption{Reflection coefficient for $m=0.4$ and wave-vector of the
  incident particle $k_0=2$ as a function of the potential barrier
  height $\phi$. This Figure is published in Ref. \cite{bisio2013dirac}.}\label{fig:Section}
\end{figure}
\begin{figure}[ht!]
\includegraphics[width=.4\textwidth]{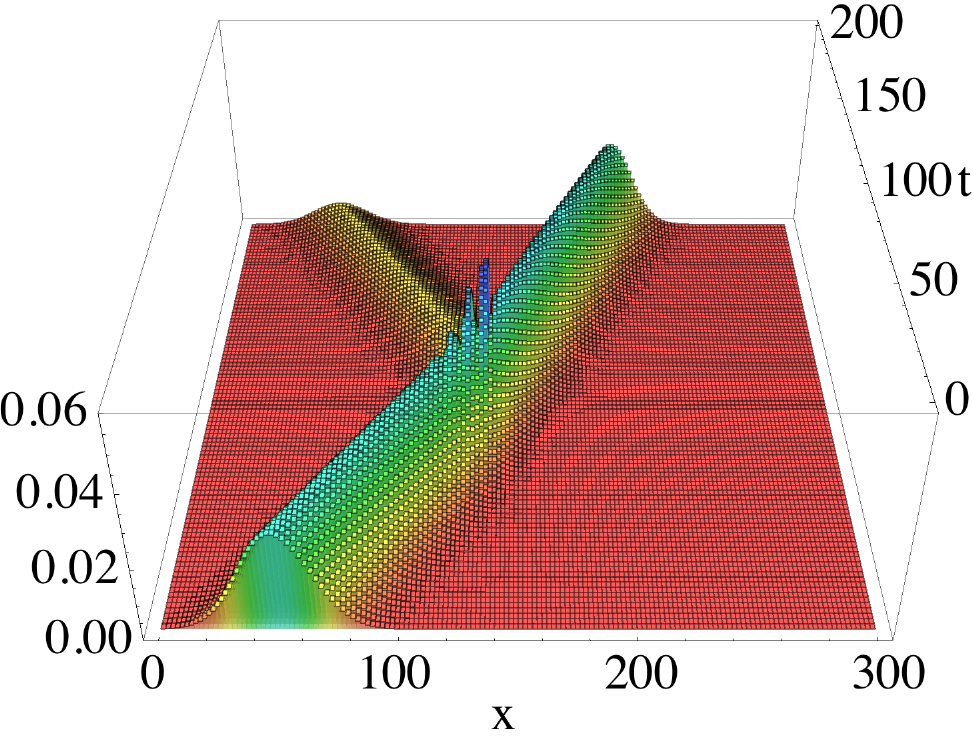}\qquad
\includegraphics[width=.4\textwidth]{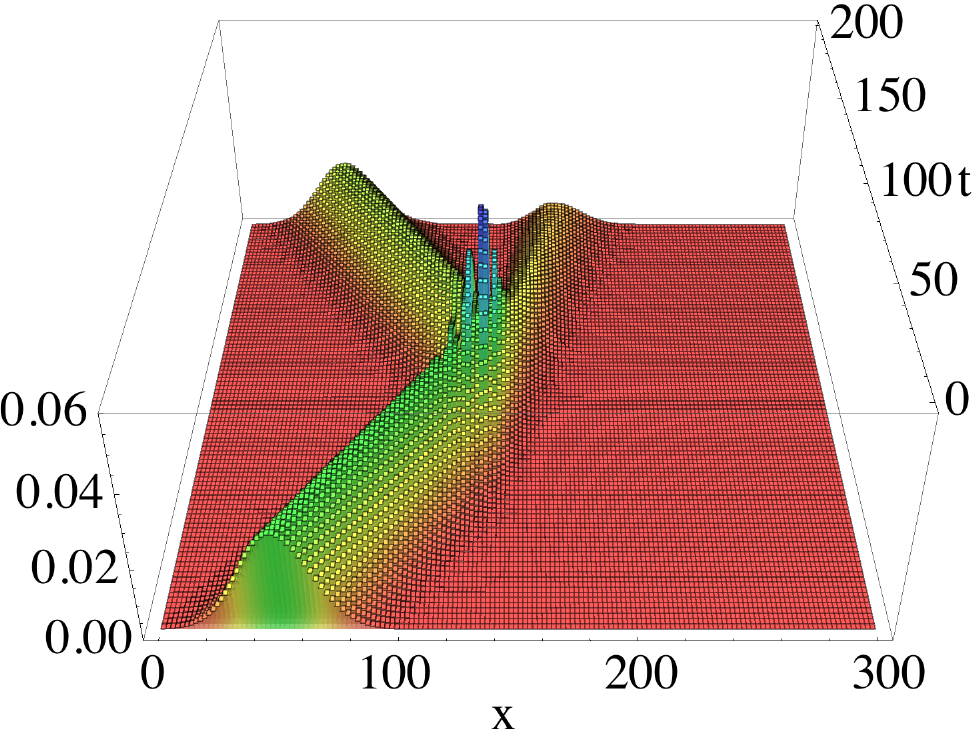}\\
\includegraphics[width=.4\textwidth]{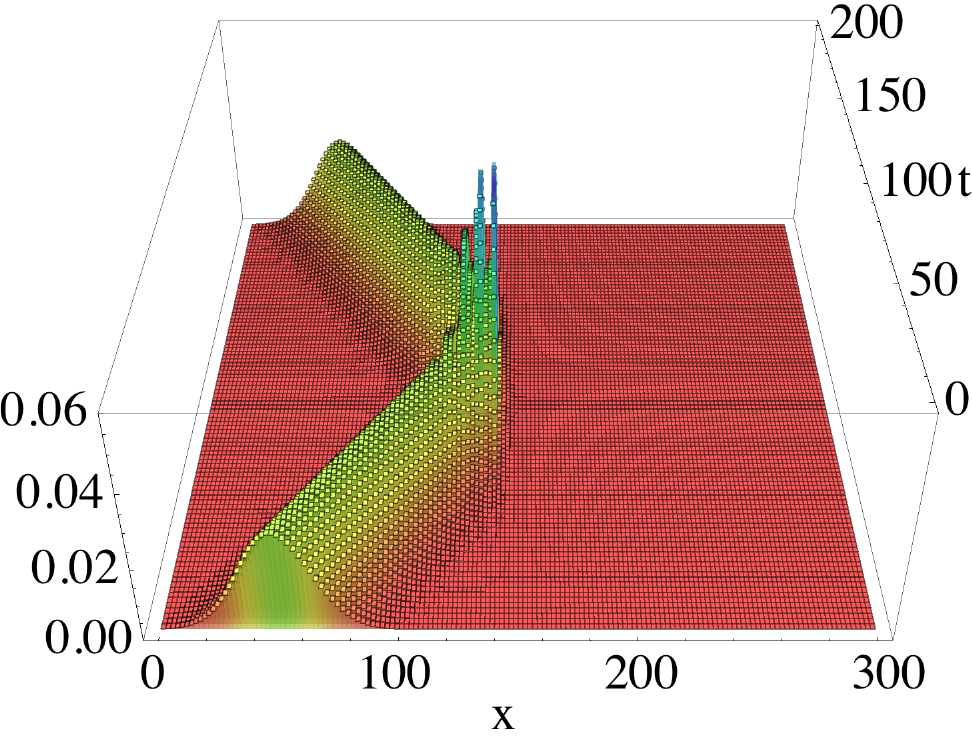}\qquad
\includegraphics[width=.4\textwidth]{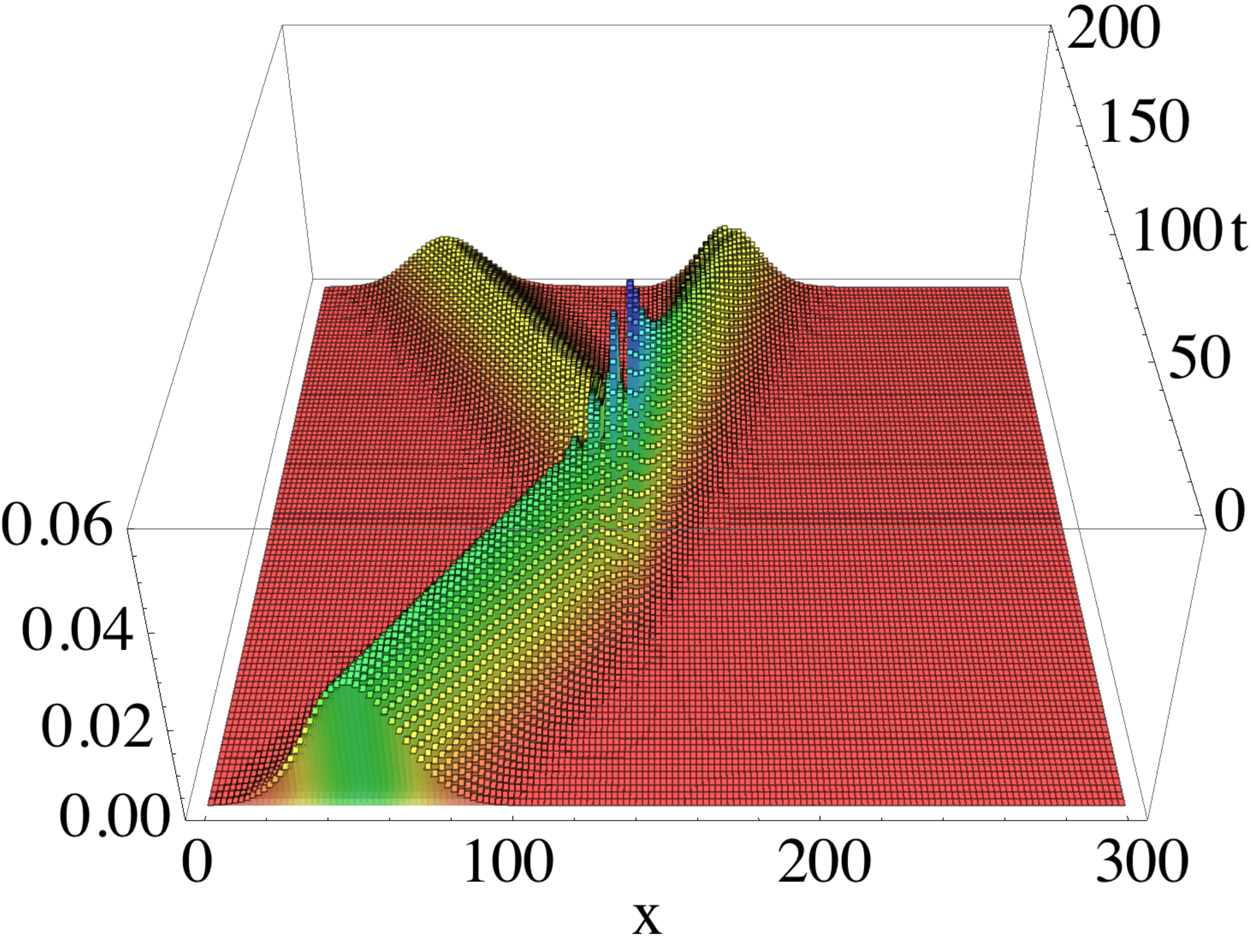}
\caption{Simulations of the one dimensional Dirac automaton evolution with a square
  potential barrier. Here the automaton mass is $m=0.2$ while the
  barrier turns on at $x=140$. In the simulation the incident state
  is a smooth state of the form $\ket{\psi(0)}=  \int
  \df{k}{\sqrt{2\pi}} g_{k_0}(k) \ket{+}_k$ peaked around the positive frequency eigenstate
  $\ket{+}_{k_0}$ with $k_0=2$ and
  with $g_{k_0}$ a Gaussian having width
  $\sigma=15^{-1}$. The incident group velocity is $v(k_0)=0.90$. The
  simulation is run for four increasing values of the potential $\phi$. 
  {\bf  Top-Left:} 
 Potential barrier height $\phi=1.42$, reflection coefficient
  $R=0.25$, velocity of the transmitted particle $v(k^\prime_0)=0.63$.  
  {\bf Top-Right:} $\phi=1.55$, $R=0.75$, $v(k^\prime_0)=0.1$.
  {\bf Bottom-Left:} $\phi=2$, $R=0.1$, $v(k^\prime_0)=0$.  
  {\bf Bottom-Right:} $\phi=2.4$, $R=0.50$, $v(k^\prime_0)=0.33$. This Figure is published in Ref. \cite{bisio2013dirac}.} \label{fig:Simulation}
\end{figure}

In Fig. \ref{fig:Section} we plot the reflection $R$ coefficient and the transmitted wave velocity
group $v(k_0')$ as a function of the potential barrier height $\phi$ with the incident wave packet
having $k_0=2$ and $m=0.4$. From the figure it is clear that after a plateau with $R=1$ the
reflection coefficient starts decreasing for higher potentials. In Fig. \ref{fig:Simulation} we show
the scattering simulation for four increasing values of the potential, say
$\phi=1.42,\,1.55,\,2,\,2.4$.

\subsection{Travel-time and Ultra-high energy cosmic rays} 

The approximated evolution studied in
Section~\ref{sec:interp-hamilt-diff}
provide a useful analytic tool for evaluating the
macroscopic evolution of the automaton.
We now consider an elementary experiment,
based on
particle fly-time, which compares
the Dirac automaton evolution with the one given by the Dirac equation.

Consider a protonwith $m_p\approx 10^{-19}$ and wave-vector peaked around
$k_{CR}\approx 10^{-8}$ in Planck units\footnote{As for order of
  magnitude, we consider numerical values corresponding to ultra high
  energy cosmic rays (UHECR) \cite{takeda1998extension}} , with a spread 
$\sigma$ of the wave-vector. We ask what is the minimal time $t_{CR}$ for observing a
complete spatial separation between the trajectory predicted by the
cellular automaton model and the one described by the usual Dirac
equation. Thus we require the separation between the two trajectories
to be greater than $\hat\sigma=\sigma^{-1}$ the initial proton's width
in the position space. We approximate the state evolution of the
wave-packet of the proton using the differential equation
\eqref{eq:schroding2} for an initial Gaussian state. The time required to
have a separation $\hat\sigma$ between the automaton and the Dirac
particle
is 
\begin{align}\label{eq:flying-time2}
t_{CR}\approx 6\frac{\hat\sigma}{m_p^2}.
\end{align}
and for $\hat\sigma=10^2\text{fm}$ (that is reasonable for a proton wave-packet) the flying time request for complete
separation between the two trajectories is
$
t_{CR}\approx 6\times 10^{60}$ Planck times, i.e. $\approx 10^{17}s$,
a value that is comparable with the age of the universe and then 
incompatible with a realistic setup.

\subsection{Phenomenology of the QCA Theory of Light} 
\label{sec:phen-qca-theory}
In this section we present an overview of the new phenomenology
emerging
 from QCA theory of free electrodynamics presented in Section
 \ref{sec:weyl-dirac-maxwell}. For a more detailed presentation we
 refer to Ref. \cite{bisio2014quantum}.

\subsubsection{Frequency dependent speed of light}
From Eq.~\eqref{eq:maxwell2} one has that
the angular frequency of the electromagnetic waves 
is given by the modified dispersion relation
\begin{align}
\label{eq:modifieddisprelmax}
\omega(\bk) = 2 | \bvec{n}_{\tfrac{\bk}{2}} |   .
\end{align}
and the usual relation $\omega(\bk) =  | \bk  |  $
is recovered in only the $| \bk  | \ll 1$ regime.
The speed of light is the group velocity of the electromagnetic 
waves, i.e.~the gradient of the dispersion relation. The major consequence 
of Eq. \eqref{eq:modifieddisprelmax} is that the speed of light depends on 
the value of $\bk$, as if the vacuum were a dispersive medium. 

The phenomenon of a $\bk$-dependent speed of light is also studied in the
quantum gravity literature where many authors considered the hypothesis that the existence of an invariant
length (the Planck scale) could manifest itself in terms of dispersion
relations that differ from the usual relativistic one
\cite{ellis1992string,lukierski1995classical,Quantidischooft1996,amelino2001testable,PhysRevLett.88.190403}.
In these models the $\bk$-dependent speed of light $c(\bk)$, at the leading order in $k :=| \bk |$,
is expanded as $c(\bk) \approx 1 \pm \xi k^{\alpha}$, where $\xi $ is a numerical factor of order
$1$, while $\alpha$ is an integer.  This is exactly what happens in our framework, where the
intrinsic discreteness of the quantum cellular automata leads to the dispersion relation of
Eq. \eqref{eq:modifieddisprelmax} from which the following $\bk$-dependent speed of light
\begin{align} \label{eq:freqdepsol}
  c(\bk) \approx 1 \pm 3\frac{k_x k_y k_z}{|\bk|^2} \approx
 1 \pm \tfrac{1}{\sqrt{3}}k,
\end{align}
can be obtained by computing the modulus of the group velocity and power expanding in $\bk$ with the
assumption $ k_x = k_y = k_z = \tfrac{1}{\sqrt{3}} k $, $k =
|\bk|$. The $\pm$ sign in Eq.~\eqref{eq:freqdepsol} depends on whether
we considered the $A^{+}(\bk)$ or the $A^{-}(\bk)$
Weyl QCA. This prediction can possibly be experimentally tested
in the astrophysical domain, where tiny corrections are
magnified by the huge time of flight.  For example, observations of the arrival times of pulses
originated at cosmological distances, like in some $\gamma$-ray
bursts\cite{amelino1998tests,abdo2009limit,vasileiou2013constraints,amelino2009prospects}, are now
approaching a sufficient sensitivity to detect corrections to the relativistic dispersion relation
of the same order as in Eq. \eqref{eq:freqdepsol}.

\begin{figure}[ht]
  \begin{center}
    \includegraphics[width=.55\textwidth]{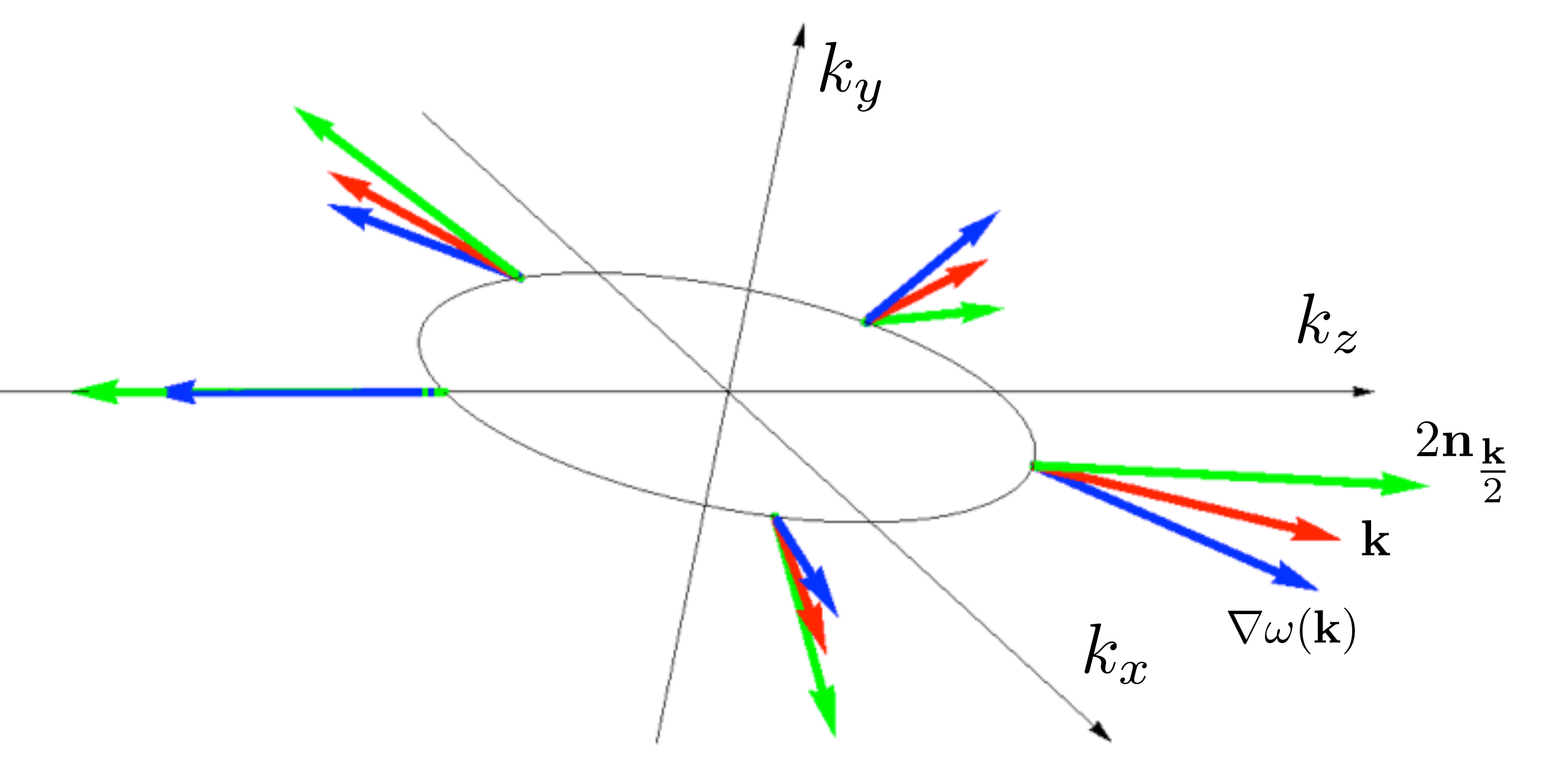} \,
    \includegraphics[width=.42\textwidth]{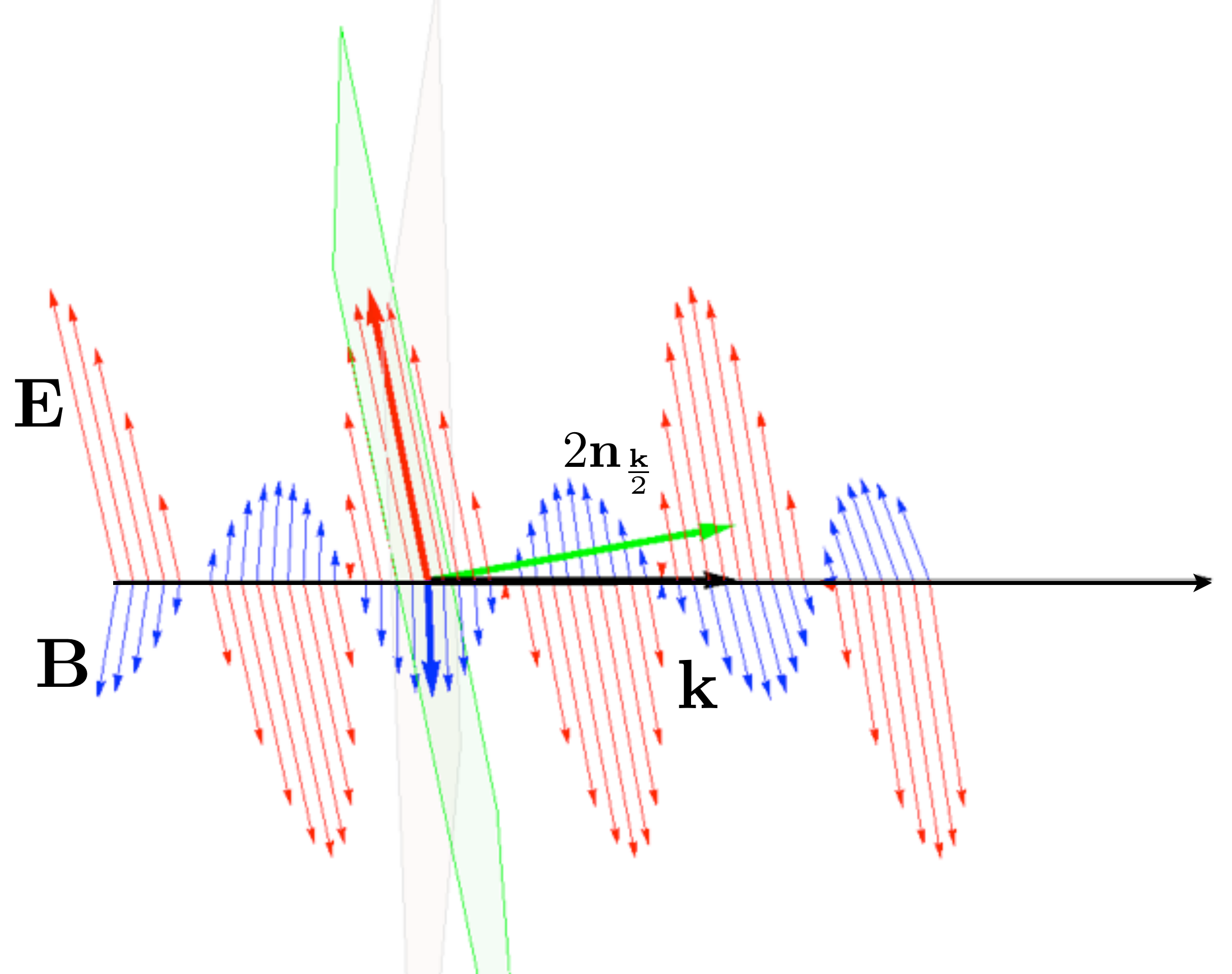}
    \caption{(colors online) {\bf Left:} the graphics shows the vector $2\bvec{n}_{\tfrac{\bk}{2}}$ (in green),
      which is orthogonal to the polarization plane, the wavevector $\bk$ (in red) and the group
      velocity $\nabla \omega (\bk)$ (in blue) as function of $\bk$ for the value $|\bk|= 0.8$ and
      different directions.
{\bf Right: }A
      rectilinear polarized electromagnetic wave. We notice that the
      polarization plane (in green) is sligtly tilted with respect the
      plane orthogonal to $\bk$ (in gray). This Figure is published in
    Ref. \cite{bisio2014quantum}.}
    \label{fig:relvectors}
  \end{center}
\end{figure}

\subsubsection{Longitudinal polarization}
A second distinguishing feature of Eq.~\eqref{eq:maxwell2} is that the polarization plane is neither
orthogonal to the wavevector, nor to the group velocity, which means that the electromagnetic waves
are no longer exactly transverse (see Fig. \ref{fig:relvectors}).  
The angle $\theta$ between the polarization plane and the plane orthogonal to $\bk$ or
$\nabla\omega(\bk)$ is of the order $\theta \approx 10^{-15}\mathrm{rad}$ for a
$\gamma$-ray wavelength, a precision which is not reachable by the present technology. Since for a
fixed $\bk$ the polarization plane is constant, exploiting greater distances and longer times does
not help in magnifying this deviation from the usual electromagnetic theory.

\subsubsection{Composite photons and modified commuation relations }

Finally, the third phenomenological consequence of the QCA theory of light is
the deviation from the exact Bosonic statistics due to the composite
nature of the photon.  As shown in Ref.\cite{bisio2014quantum},
the choice of the function ${f}_\bk(\bvec{q})$ in Eq.~\eqref{eq:electric and magnetic field}
determines the regime where the composite photon can be approximately treated as a Boson.  However,
independently on the details of function ${f}_\bk(\bvec{q})$, one can prove that a Fermionic
saturation of the Boson is not visible, e.g. for the most powerful laser \cite{dunne2007high} one
has a approximately an Avogadro number of photons in $10^{-15}$cm${}^3$, whereas in the same volume
on has around $10^{90}$ Fermionic modes.
Another test for the composite nature of photons is provided by the prediction of
deviations from the Planck's distribution in Blackbody
radiation experiments. A similar analysis was carried out in
Ref. \cite{perkins2002quasibosons}, where the author showed that the predicted 
deviation from Planck's law is less than one part over $10^{-8}$, 
well beyond the sensitivity of present day experiments.

\section{Future perspectives}

We conclude this paper with an overview of the future developments of
the research program on QCA for Field Theory.

\subsection{Lorentz covariance and Deformed Relativity}
\label{sec:lorentz-covar-deform}
Because of the intrinsic discreteness of the model, a dynamical evolution described in terms of a QCA cannot satisfy the usual Lorentz covariance, which must break down at the Planck scale. Moreover the very notions of spacetime and boosted reference frame break down at small scales, and need a thoughtful reconsideration. In Ref.~\cite{bisio2015lorentz} a definition of reference frame was introduced in a background-free scenario, in terms of labelling of irreducible representations of the group $G$. The Lorentz symmetry is then recovered by imposing a generalized relativity principle on possible changes of reference frame, allowing only those changes that leave the automaton invariant. A preliminary analysis of the one-dimensional case can be found in Ref.~\cite{bibeau2013doubly}, where only the necessary condition of preserving the dispersion relation was considered. Focusing on the one dimensional Dirac QCA we have
\begin{align}
  \label{eq:dispreldirac1d}
  \omega(k) = \arccos(\sqrt{1-m^2} \cos(k))
\end{align}
and one can see that in the $k \ll 1$,$m \ll 1$ limit Eq.~\eqref{eq:dispreldirac1d} reduces to the usual relativistic
dispersion relation $\omega^2=k^2+m^2$. It is also immediate to check that the automaton dispersion relation of
Eq.~\eqref{eq:dispreldirac1d} is not invariant under standard Lorentz transformation.  In order to preserve Eq.~\eqref{eq:dispreldirac1d} one needs to introduce a non-linear representation of the Lorentz transformation in the wave-vector space---as proposed in the so
called \emph{deformed special relativity} (DSR) models
\cite{amelino2002relativity,amelino2001planck,amelino2002quantum,amelino2001testable,PhysRevLett.88.190403,magueijo2003generalized}.

In Ref.~\cite{bisio2015lorentz} the boosts preserving the three-dimensional Weyl automaton were then derived in the form of the following non-linear representation of the Lorentz group
\begin{equation}\label{eq:non-linear-boosts}
  \B:= \M^{-1}\circ\L\circ\M,
\end{equation}
where $\M:\Reals^4\to\Reals^4$ is a non-linear map. 
The specific form of $\M$ gives rise to a particular frequency/wave-vector Lorentz deformation. 

These ideas can also be applied to the three-dimensional Dirac QCA. In this case one can show that a change of the rest mass should be involved in the representation of boosts, in order to obey our generalized relativity principle. Interestingly, this unexpected feature gives rise to an emergent space-time with a non-linear de Sitter symmetry instead of the Lorentz one.

Another challenging line of research is to characterize the emergent spacetime of the QCA framework. The DSR models provide a complete description of Lorentz symmetry in frequency/wave-vector space but there are heuristic 
ways to extend this framework to the position-time space. Relative locality \cite{amelino2011relative,amelino2013relative},
non-commutative spacetime \cite{connes1991particle} and Hopf algebra symmetries \cite{lukierski1991q,majid1994bicrossproduct}  
have been considered in order to give a real space formulation of deformed relativity.

Finally, we would like to stress that space-time emerges from: i) the structure of the group $G$, ii) the specific expression of the automaton and iii) the generalized relativity principle, while all these concepts do not require any space-time background. Thus, outside the limits in which the relativistic approximations hold, the very structure of our usual space-time break down, substituted by other counterintuitive effects. In particular this is true in all physical situations where the discrete structure of the lattice $G$ becomes relevant.

\subsection{Thermodynamics of free ultra-relativistic particles and QCA}

\label{sec:therm-free-ultr}
Most of the analysis that we presented in this paper was focused on
the dynamics of one particle state and the deviations of this
kinematics from the usual relativistic one. On the other hand, it would be
interesting to explore the QCA phenomenology when the number of
particle goes to infinity, namely a thermodynamic limit. Since the QCA
we are considering descrive a non-interacting dynamics, the
thermodynamic that will emerge will describe a gas of free particles. 
However, since the dispersion relation of the QCA differs from the
relativistic one, the density of states will be different. 

In the case
of free fermions this will result in a shift of the Fermi energy
that could become relevant when the number of fermions becomes very
large. One could for example analyze how the
Chandrasekhar limit of white dwarfs is modified in this context (see
Ref. \cite{amelino2012uv,camacho2006white} for a similar analysis in a
different context).

\subsection{Interacting extensions}

The theory of linear QCAs naturally leads to free quantum field theories. In order to introduce interactions, one needs to relax the linearity assumption. This can be done by splitting the computational step in two stages, the first one acting linearly, and the second one representing a nonlinear and completely local evolution. This can be motivated in terms of a time-local gauge symmetry that must preserve some local degree of freedom, in particular the local number of excitations. This simple modification of the linear automaton introduces a non-trivial interaction, making the automaton non-trivially reducible to a quantum walk. Preliminary analysis shows that this minimal relaxation of linearity is sufficient to give rise to couplings that might reproduce the phenomenology of quantum electrodynamics.

\section*{Acknowledgements}
This work has been supported in part by the Templeton Foundation under the project ID\# 43796 {\em A
  Quantum-Digital Universe}.

\bibliographystyle{apsrev4-1}
\bibliography{bibliography}

\end{document}